\documentclass[12pt]{article}
\usepackage{a4wide,epsfig,amsmath,amssymb,cite,scalefnt}
\newcommand{\dd}{{\rm d}}
\newcommand{\order}[1]{{\cal O}\!\left(#1\right)}
\newcommand{\eqn}[1]{Eq.\,(\ref{#1})}

\newcommand{\lM}{L_{\tilde M}}

\newcommand{\ltM}{L_{t\tilde M}}

\newcommand{\mudec}{\mu_{\rm dec}}
\newcommand{\mugut}{\mu_{\abbrev\rm GUT}}

\newcommand{\abbrev}{\scalefont{.9}}
\newcommand{\drbar}{$\overline{\mbox{\abbrev DR}}$}
\newcommand{\msbar}{$\overline{\mbox{\abbrev MS}}$}
\newcommand{\drbarmath}{\overline{\rm\abbrev DR}}
\newcommand{\msbarmath}{\overline{\rm\abbrev MS}}
\newcommand{\mbDR}[1]{m_b^{\drbarmath{}#1}}

\newcommand{\asDR}[1]{\alpha_s^{\drbarmath{}#1}}
\newcommand{\asMS}[1]{\alpha_s^{\msbarmath{}#1}}

\newcommand{\qcd}{{\abbrev QCD}}
\newcommand{\susy}{{\small SUSY}}
\newcommand{\dreg}{{\small DREG}}
\newcommand{\dred}{{\small DRED}}

\newcommand{\mDRbar}{m^{\overline{\rm DR}}}
\newcommand{\mMSbar}{m^{\overline{\rm MS}}}

\begin{document}



\title{\vskip-3cm{\baselineskip14pt
    \begin{flushleft}
      \normalsize SFB/CPP-07-31 \\
      \normalsize TTP/07-12 \\
       \normalsize WUB/07-04 
  \end{flushleft}}
  \vskip1.5cm
  Running of $\alpha_s$ and $m_b$ in the MSSM
}
\author{\small R.V. Harlander$^{(a)}$,
  L. Mihaila$^{(b)}$, M. Steinhauser$^{(b)}$\\
  {\small\it (a) Fachbereich C, Theoretische Physik,
    Universit{\"a}t Wuppertal,}\\
  {\small\it 42097 Wuppertal, Germany}\\
  {\small\it (b) Institut f{\"u}r Theoretische Teilchenphysik,
    Universit{\"a}t Karlsruhe,}\\
  {\small\it 76128 Karlsruhe, Germany}\\
}

\date{}

\maketitle

\thispagestyle{empty}

\begin{abstract}
A consistent evolution of the strong coupling constant $\alpha_s$ from
$M_Z$ to the GUT scale is presented, involving three-loop running and
two-loop decoupling. The two-loop transition from the \msbar{}- to the
\drbar{}-scheme is properly taken into account.  In the second part of
the paper, the bottom quark mass in the $\overline{\mbox{ DR}}$- scheme at the
electroweak/SUSY scale is evaluated with four-loop accuracy.
We find that the three-loop effects are comparable to the experimental
uncertainty both for $\alpha_s$ and $m_b$.

\medskip


\end{abstract}


\section{Introduction}

The Minimal Supersymmetric Standard Model (MSSM) is one of the most
studied extensions of the Standard Model (SM).  The observation that the
gauge couplings of the strong, electromagnetic and weak interaction tend
to unify in the MSSM at a high energy scale $\mugut{}\simeq
10^{16}$~GeV~\cite{Dimopoulos:1981yj,Ibanez:1981yh,Amaldi:1991cn,Langacker:1991an,Ellis:1990wk}
consistent with Grand Unification Theories (GUT) is probably the most
compelling hint in favour of supersymmetry (SUSY). Constraints from the
Yukawa sectors of SUSY-GUT models yield appealing predictions for SM
parameters, including the top quark mass and the ratio of the bottom
quark to the tau lepton
masses~\cite{Arason:1991hu,Barger:1992ac,Carena:1994bv,Hall:1993gn}.

Nevertheless, SUSY can only be an approximate symmetry in
nature. Several scenarios for the mechanism of SUSY breaking, based on
very different physical models, have been proposed. A possibility to
constrain the type and scale of SUSY breaking is to study, with very
high precision, the relations between the MSSM parameters evaluated at
the electroweak and the GUT scales.  The extrapolation over many orders
of magnitude requires high-precision experimental data at the low energy
scale. A first set of precision measurements is expected from the CERN
Large Hadron Collider (LHC) with an accuracy at the percent
level. A comprehensive high-precision analysis can be performed at the
International Linear Collider (ILC), for which the
estimated experimental accuracy is at the per mill level. It is
obvious that the same precision must be reached
also on the theory side in order to match with the
data~\cite{Aguilar-Saavedra:2005pw}. Running analyses based on full
two-loop renormalization group equations
(RGEs)~\cite{Martin:1993yx,Martin:1993zk,Jack:1994kd,Yamada:1994id} for
all parameters and one-loop threshold corrections~\cite{Pierce:1996zz}
are currently implemented in the public programs
ISAJET~\cite{Paige:2003mg}, SOFTSUSY~\cite{Allanach:2001kg},
SPHENO~\cite{Porod:2003um}, SuSpect~\cite{Djouadi:2002ze}. The agreement
between the different codes is in general within one
percent~\cite{Allanach:2003jw}. A first three-loop running analysis,
based, however, only on one-loop threshold effects, was carried out in
Ref.~\cite{Ferreira:1996ug,Jack:2003sx}. While the three-loop running
has little effect on the weakly interacting particle sector of the MSSM,
this effect can be comparable to that of two loop running for strongly
interacting particles, as for example the squark mass
spectrum~\cite{Jack:2003sx}.
 
In this paper, we elaborate an earlier
investigation~\cite{Harlander:2005wm} of the evolution of the
strong coupling $\alpha_s$  in  MSSM, based on 
three-loop RGEs~\cite{Jack:1996vg} and two-loop threshold corrections. 
On the one hand, the  three-loop corrections
reduce  significantly the dependence on the scale at which heavy
particles are integrated out  and on the other hand, they are
essential for  phenomenological studies, because they are as large as, or
greater than, the effects induced by the current experimental accuracy of
$\alpha_s(M_Z)$. 
Our aim is to compute $\alpha_s$  at a high-energy scale
$\mu\simeq{\cal O}(\mugut{}) $ with three-loop accuracy, starting from the
SM parameter $\alpha_s(M_Z)$ as input. 
Additionally, we compare the predictions obtained within the above
mentioned approach with those based on the
leading-logarithmic (LL) approximation suggested in
Ref.~\cite{Aguilar-Saavedra:2005pw}. 

In the context of $SO(10)$ GUT~\cite{Hall:1993gn,Rattazzi:1995gk} the
third-family Yukawa couplings may unify for a large ratio of the Higgs
vacuum expectation values $\tan\beta \simeq 50$. Moreover, in all SUSY
models with large $\tan\beta$, the supersymmetric particle spectrum and
the Higgs boson masses are sensitive to the bottom Yukawa
coupling~\cite{Baer:1997yi,Baer:1998sz}.  In turn, the relation between
the running bottom mass and the bottom Yukawa coupling is affected in
such theories by large supersymmetric radiative
corrections~\cite{Carena:1999py}. For the calculation of higher order
corrections in the framework of SUSY theories, the \drbar{} scheme,
based on regularization by dimensional reduction (DRED) and modified minimal
subtraction, is most convenient.  On the other hand, the values of the
SM parameters are presently extracted from the experimental data for the
\msbar{} renormalization scheme. It is obvious that translation formulae
from one scheme to the other to the appropriate order in
the perturbative expansion have to be employed for phenomenological studies.

Owing to these considerations, a precise determination of the running bottom
mass in the \drbar{} scheme is of great importance for SUSY theories
sensitive to the bottom quark Yukawa coupling.  Ref.~\cite{Baer:2002ek}
proposed an approach to the determination of $m_b^{\drbarmath}(M_Z)$, which
avoids the infrared sensitivity problem related with the use of the
bottom-quark pole mass.  The procedure consists of employing the RGEs of
QCD and the relation between {\msbar} and {\drbar} masses in order to
calculate $m_b^{\drbarmath}$, using as input parameter the
accurately known $\mu_b= m_b^{\msbarmath}(\mu_b)$.  A second purpose of
this paper is to extend the study of Ref.~\cite{Baer:2002ek} to three-
and even to four-loop order~\cite{Harlander:2006rj,Harlander:2006xq}, in
order to take full advantage of the recent determination of
$m_b^{\msbarmath}(\mu_b)$ with four-loop
accuracy~\cite{Kuhn:2007vp}. Furthermore, we study the phenomenological
significance of evanescent couplings related with the application of
the \drbar{} scheme to QCD.

The remainder of the paper is organized as follows: in
Section~\ref{sec::as} we discuss the evaluation of the strong coupling
at the GUT scale from the knowledge of $\alpha_s(M_Z)$ and propose a
consistent prescription based on three-loop running and two-loop
decoupling. 
Section~\ref{sec::mb} deals with the running of the bottom quark mass
from $\mu=\mu_b$ to a decoupling scale of supersymmetric particles.


\section{\label{sec::as}Running and decoupling of \boldmath{$\alpha_s$}}

The value of the strong coupling constant, measured at the mass of the
$Z$ boson $M_Z$, is a central quantity in high energy physics.  Usually,
what is being quoted in the literature as $\alpha_s(M_Z)$ is its value
in a theory with five quark flavours, renormalized in the \msbar{}
scheme. In order to be precise, we will refer to this quantity as
$\asMS{,(5)}(M_Z)$.

On the contrary, what is meant by the value of $\alpha_s(\mugut)$ is the
strong coupling constant in a supersymmetric theory, renormalized in the
\drbar{} scheme. We will denote this quantity by $\asDR{,(\rm
full)}(\mugut)$ in what follows.  It is the purpose of this section to
study the relation between $\asMS{,(5)}(M_Z)$ and $\asDR{,(\rm
full)}(\mugut)$ at three-loop level. This requires the consistent
combination of (a) the renormalization group evolution of $\alpha_s$;
(b) the transition from the \msbar{} to the \drbar{} scheme; (c) the
transition from five-flavour QCD to the full SUSY theory. All these
issues will be discussed in detail in what follows.

\subsection{Renormalization group evolution}

The energy dependence of the strong coupling constant 
is governed by the RGE
\begin{equation}
\begin{split}
\mu^2\frac{\dd}{\dd\mu^2} \alpha_s(\mu^2) = \beta(\alpha_s)\,,\qquad
\beta(\alpha_s) = -\alpha_s^2\sum_{n\geq 0}\beta_n\alpha_s^n\,.
\label{eq::betadef}
\end{split}
\end{equation}
The $\beta$ function depends on the underlying theory and on the
renormalization scheme. In QCD with $n_f$ quark flavours, it is known in
the \msbar{} scheme through four
loops~\cite{vanRitbergen:1997va,Czakon:2004bu}; the 
first three coefficients read
\begin{equation}
\begin{split}
\beta_0 &= \frac{1}{4}\left(11 - \frac{2}{3}n_f\right)\,,\qquad
\beta_1 = \frac{1}{16}\left( 102 - \frac{38}{3}\,n_f \right)\,,\\
\beta_2 &= \frac{1}{64}\left(
\frac{2857}{2} - \frac{5033}{18}\,n_f + \frac{325}{54}n_f^2\right)\,.
\label{eq::betaqcd}
\end{split}
\end{equation}
The QCD $\beta$ function in the \drbar{} scheme involves evanescent
couplings (see Section~\ref{sec::msdr} below)
at three loops and
higher. Explicit results through four loops are given in
Refs.\,\cite{Harlander:2006rj,Harlander:2006xq}.

In SUSY-QCD, the $\beta$ function has been evaluated in the \drbar{}
scheme through three loops from the NSVZ
formula~\cite{Novikov:1985rd,Shifman:1986zi} by a proper shift in the
strong coupling constant~\cite{Jack:1996vg}. For six (s)quark flavours
one finds
\begin{equation}
\begin{split}
\beta_0 &= \frac{3}{4}\,,\qquad
\beta_1 = -\frac{7}{8}\,,\qquad
\beta_2 = -\frac{347}{192}\,.
\label{eq::betasusy}
\end{split}
\end{equation}

\subsection{Decoupling of the SUSY particles}

For mass independent renormalization schemes like \msbar{} or \drbar{},
the decoupling of heavy particles has to be performed explicitely.  In
practice, this means that intermediate effective theories are introduced
by integrating out the heavy degrees of freedom. The parameters of the
various effective theories are related by so-called decoupling (or matching)
constants.

When going from a high to a low energy scale, one may separately
integrate out every particle at its individual threshold. This
multi-scale approach (sometimes denoted as ``step approximation'') is
suited for SUSY models with a severely split mass
spectrum~\cite{Castano:1993ri}. As pointed out long time ago, the
intermediate effective theories with ``smaller'' symmetry raise the
problem of introducing new couplings, each governed by its own RGE. This
is the case in the Yukawa sector if the $SU(2)$ symmetry is broken, for
example. It also happens for the ``evanescent'' couplings, related with
the application of \dred{} to non-supersymmetric theories. A
comprehensive multi-scale decoupling procedure in the context of the MSSM
is currently not available, although it could be of phenomenological
relevance~\cite{Baer:2005pv}.

On the other hand, in SUSY models with roughly degenerate mass spectrum
at a scale $\tilde M$, one can consider the MSSM as the full theory that
is valid from the GUT scale $\mugut{}$ down to $\tilde{M}$, which we
assume to be around $1$\,TeV. Integrating out all SUSY particles at this
common scale, one directly obtains the SM as the effective theory, valid
at low energies.  As already mentioned above, the strong coupling
constants in SUSY and in QCD are then related by a decoupling constant
$\zeta_s$. The transition between the two theories can be done at an
arbitrary decoupling scale $\mu$:\footnote{ In principle the decoupling
constants should also carry a label \msbar{} or \drbar{}. However, in
this paper decoupling is always performed in the \drbar{} scheme, and
therefore we omit the label. }
\begin{equation}
\begin{split}
\asDR{,(n_f)}(\mu) &= \zeta_s^{(n_f)}\,\asDR{,(\rm
  full)}(\mu)\,.
\label{eq::asdec}
\end{split}
\end{equation}
$\zeta_s$ depends logarithmically on the scale $\mu$, which is why one
generally chooses $\mu\sim \tilde M$.  In \eqn{eq::asdec}, $n_f=6$ means
that only the SUSY particles are integrated out, while for $n_f=5$ at
the same time the top quark is integrated out.  This procedure, also
known as ``common scale approach''~\cite{Baer:2005pv}, has the advantage
that it avoids the occurrence of intermediate non-supersymmetric
effective theories.  Most of the present codes computing the SUSY
spectrum~\cite{Porod:2003um,Allanach:2001kg,Djouadi:2002ze} follow this
approach by applying the one-loop approximation of \eqn{eq::asdec} and
setting $n_f=5$ and $\mu=M_Z$.

In this paper, we apply the corresponding two-loop approximation, as it
is required by a consistent three-loop evolution of the coupling
constant~\cite{Hall:1980kf}.  Furthermore, in contrast to most analyses
performed up to now, we allow for a general decoupling scale $\mu=\mudec$
where the heavy particles are decoupled.

The decoupling coefficients of \eqn{eq::asdec} have been evaluated
through two loops in Eqs.\,(26) and (29) of
Ref.\,\cite{Harlander:2005wm}:
\begin{equation}
\begin{split}
  \zeta_s^{(n_f)} &= 1 
  + \frac{\alpha_s}{\pi}\,\zeta_{s1}^{(n_f)}
  + \left(\frac{\alpha_s}{\pi}\right)^2\,\zeta_{s2}^{(n_f)}
  + \order{\alpha_s^3}\,,
\end{split}
\end{equation}
where $\alpha_s \equiv \asDR{,({\rm full})}(\mu)$, and
\begin{equation}
\begin{split}
   \zeta_{s1}^{(6)} &= - \lM\,,\qquad
  \zeta_{s1}^{(5)} = -
    \frac{1}{6} L_t
    - L_{\tilde{M}}\,,\\
   \zeta_{s2}^{(6)} &=
    - \frac{65}{32}
      - \frac{5}{2} L_{\tilde{M}} 
      + \lM^2
    \,,\\
   \zeta_{s2}^{(5)} &=
    - \frac{215}{96}
      - \frac{19}{24} L_t
      - \frac{5}{2} L_{\tilde{M}} 
      + \left[
      \frac{1}{6} L_t
      + L_{\tilde{M}}\right]^2
      \\&\qquad
      + \left(\frac{m_t}{\tilde M}\right)^2 \left(  
       \frac{5}{48} 
      + \frac{3}{8}\ltM 
      \right)
      -  \frac{7\pi}{36} \left(\frac{m_t}{\tilde M}\right)^3
      \\&\qquad
      + \left(\frac{m_t}{\tilde M}\right)^4 \left(  
      \frac{881}{7200} 
      - \frac{1}{80}\ltM 
      \right)
      + \frac{7\pi}{288} \left(\frac{m_t}{\tilde M}\right)^5
      + \ldots
  \,.
\label{eq::zetas}
\end{split}
\end{equation}
$\tilde M$ is the common mass of the \susy{} particles, $m_t$ is the top
quark mass, and
\begin{equation}
\begin{split}
\lM &= \ln\frac{\mu^2}{\tilde M^2}\,,\qquad
L_t = \ln\frac{\mu^2}{m_t^2}\,,\qquad
\ltM = \ln\frac{m_t^2}{\tilde M^2}\,.
\end{split}
\end{equation}
Although the formula for $\zeta_{s2}^{(5)}$ in \eqn{eq::zetas} has been
obtained in the limit $m_t\ll \tilde M$, it has been shown that it holds at
the per cent level for all values of $m_t\leq \tilde
M$~\cite{Harlander:2005wm}. The exact limit $m_t=\tilde M$ is also given
in Ref.\,\cite{Harlander:2005wm}.

\subsection{\msbar{}--\drbar{} conversion and evanescent
  couplings}\label{sec::msdr}

In order to use \eqn{eq::asdec}, one needs to transform the input value
for $\alpha_s$ from the \msbar{} to the \drbar{} scheme. The
corresponding relation has been evaluated through three loops in
Ref.\,\cite{Harlander:2006xq}. Here, we only need the two-loop
expression~\cite{Harlander:2006rj}:
\begin{equation}
\begin{split}
  \asMS{} &= \asDR{}\left[1-\frac{\asDR{}}{4\pi}
    -    \frac{5}{4} \left(\frac{\asDR{}}{\pi}\right)^2
    + \frac{\asDR{}\alpha_e^{}}{12\pi^2}\,n_f
    + \ldots \right]
  \,,
  \label{eq::asMS2DR_2}
\end{split}
\end{equation}
where $\asDR{} \equiv \asDR{,(n_f)}(\mu)$ and $\asMS{} \equiv
\asMS{,(n_f)}(\mu)$. 

$\alpha_e{} \equiv \alpha_e^{(n_f)}(\mu)$ is
one of the so-called evanescent coupling constants that occur when
\dred{} is applied to non-supersymmetric theories (QCD in this case).
In particular, it describes the coupling of the
$2\varepsilon$-dimensional components (so-called $\varepsilon$-scalars)
of the gluon to a quark. Other evanescent couplings that occur in QCD
are $\eta_r$ ($r=1,2,3$), related to the vertex of four
$\varepsilon$-scalars (see, e.g., Ref.\,\cite{Harlander:2006xq} for
details).  In the following, we will assume that QCD is obtained by
integrating out the heavy degrees of freedom (squarks and gluinos) from
SUSY-QCD. In this case, the evanescent couplings can be related to the
gauge coupling $\alpha_s$ as follows:\footnote{Assuming that
\dred{} preserves supersymmetry.}
\begin{eqnarray}
  &&\alpha_e^{(\rm full)}(\mu) = 
  \asDR{,(\rm full)}(\mu) =
  \eta_1^{(\rm full)}(\mu)
  \,,
  \nonumber\\
  &&\eta_2^{(\rm full)}(\mu) 
  = \eta_3^{(\rm full)}(\mu) = 0
  \,.
  \label{eq::couplings_full}
\end{eqnarray}
The evanescent couplings in $n_f$-flavour QCD, i.e.\ $\alpha_e^{(n_f)}$
and $\eta_r^{(n_f)}$ are then obtained by decoupling relations analogous
to \eqn{eq::asdec}.

\subsection{Decoupling for \boldmath{$\alpha_e$}}
For the one-loop decoupling relation that connects the values of
$\alpha_e$ in five-flavour QCD and in SUSY-QCD, we find by a calculation
similar to Ref.\,\cite{Harlander:2005wm}:
\begin{eqnarray}
  \alpha_e^{(5)}(\mu) &=& \zeta_e \alpha_e^{(\rm full)}(\mu)
  \,,\nonumber\\
  \zeta_e &=& 1 + \frac{\asDR{,\rm (full)}(\mu)}{\pi}\bigg[
    T_F \left(-\frac{1}{2} L_t\right)
    + C_A \left(-\frac{1}{4}L_{\tilde{g}}
    +\frac{
      \left(m_{\tilde{g}}^2 
      (1+L_{\tilde{g}})-m_{\tilde{q}}^2 (1+L_{\tilde{q}})\right)
    }{2(m_{\tilde{g}}^2-m_{\tilde{q}}^2)}
    \right)\nonumber\\
    &&+
    C_F \left(\frac{m_{\tilde{q}}^2-3 m_{\tilde{g}}^2}{4
      (m_{\tilde{g}}^2-m_{\tilde{q}}^2)}
    +\frac{m_{\tilde{q}}^2 (2 m_{\tilde{g}}^2-m_{\tilde{q}}^2 )
      L_{\tilde{q}} - m_{\tilde{g}}^4 L_{\tilde{g}} }
    {2 (m_{\tilde{g}}^2-m_{\tilde{q}}^2)^2}
    \right)
    \bigg] + \order{\alpha_s^2}
  \,,
  \label{eq::decoupling_ev}
\end{eqnarray}
where we have used the short-hand notation 
\begin{equation}
  L_x = \ln\frac{\mu^2}{m_x^2}\,, \qquad x\in\{t,\tilde{g},\tilde{q}\}\,.
\end{equation}
Furthermore, the mixing angle between the squark mass eigenstates
$m_{\tilde{q}1}$ and $m_{\tilde{q}2}$ has been set to zero.
Note that \eqn{eq::decoupling_ev} depends on a specific (s)quark flavour
$q$.
This means that actually $\zeta_e$
and therefore also $\alpha_e$ should carry an additional label $q$:
\[
\begin{split}
  \alpha_e^{q,(5)}(\mu) &= \zeta_e^{q}\,\alpha_e^{\rm (full)}(\mu)\,.
\end{split}
\]
For example, $\alpha_e^{u,(5)}$ is the coupling constant for the $\bar u
u\varepsilon$ vertex, where the decoupling constant $\zeta_e^u$ depends
on $m_{\tilde u}$. Therefore, if all squark masses are different, one
needs a separate evanescent coupling for each quark flavour.  We remark
that gauge invariance prevents the occurrence of such a flavour
dependence for $\asDR{,(5)}$. If all squarks have the same mass
$m_{\tilde{q}}$, then all evanescent couplings are the same,
$\alpha_e^{u,(5)} = \alpha_e^{d,(5)} = \cdots = \alpha_e^{b,(5)}$, and
one can drop the flavour label.
In the limit $m_{\tilde{g}} = m_{\tilde{q}}= \tilde {M}$,
Eq.~(\ref{eq::decoupling_ev}) reads
\begin{eqnarray}
  \zeta_e &=& 1 + \frac{\asDR{,\rm (full)}(\mu)}{\pi}\bigg[
    T_F \left(-\frac{1}{2} L_t\right)
    + C_A \left(\frac{1}{4}L_{\tilde M}\right)
    + C_F\left(-\frac{1}{2}L_{\tilde M}\right)\bigg]
  \,.
\end{eqnarray}
$\alpha_e$ occurs typically at higher orders, and in fact, as we will
now argue, one can easily eliminate it perturbatively from the three-loop
running of $\alpha_s$.

Because of \eqn{eq::couplings_full}, one can write
\begin{equation}
  \begin{split}
    \alpha_e^{(n_f)}(\mu) = \xi^{(n_f)}\,\asDR{,(n_f)}(\mu)\,,
  \end{split}
\end{equation}
with
\begin{equation}
  \begin{split}
    \xi^{(n_f)} &= 
    \frac{\zeta_e^{(n_f)}}{\zeta_s^{(n_f)}} =
    1 + \frac{\asDR{,(n_f)}(\mu)}{\pi}\, \xi_1^{(n_f)}
    + {\cal O}(\alpha_s^2)
    \,,
    \label{eq::xieff}
  \end{split}
\end{equation}
where the dependence on $\asDR{,({\rm full})}$ has been eliminated by using
\eqn{eq::asdec}. 
In case all heavy particles have a common mass $\tilde M$,
the quantity $\xi_1^{(n_f)}$ only depends on $\ln(\mu^2/\tilde M^2)$ and
thus for $\mu\equiv\mu_{\rm dec}\approx \tilde M$ no numerically enhanced
perturbative coefficients appear. This is even true for a mass spectrum
where all mass ratios are within the range of one or two orders of
magnitude.  This absence of artificially large logarithms allows us to
write
\begin{equation}
  \begin{split}
    \alpha_e^{(n_f)}(\mudec) &= \asDR{,(n_f)}(\mudec) 
    + \order{\alpha_s^2} 
  \,.
    \label{eq::aemudec}
  \end{split}
\end{equation}
In this way, we can replace $\alpha_e$ by $\asDR{}$
in the two-loop term of Eq.~(\ref{eq::asMS2DR_2}).

\subsection{Evaluation of \boldmath{$\alpha_s(\mugut)$} from
  \boldmath{$\alpha_s(M_Z)$} }

There are various ways to go from $\asMS{,(n_f)}(M_Z)$ to $\asDR{,\rm
  (full)}(\mugut)$. They most importantly differ in the degree to which
  evanescent couplings are needed.  We propose the following
  method:
\begin{equation}
  \begin{split}
    \asMS{,(n_f)}(M_Z)\quad
    &\stackrel{(i)}{\to}\quad \asMS{,(n_f)}(\mudec)\quad
    \stackrel{(ii)}{\to}\quad \asDR{,(n_f)}(\mudec)\\
    &\stackrel{(iii)}{\to} \quad\asDR{,(\rm full)}(\mudec)\quad
    \stackrel{(iv)}{\to}\quad \asDR{,(\rm full)}(\mugut)\,.
  \end{split}
  \label{eq::asrundec}
\end{equation}
The individual steps require
\renewcommand{\labelenumi}{($\roman{enumi}$)}
\begin{enumerate}
\item $\beta(\alpha_s)$ in \qcd{} through three loops
  [Eqs.\,(\ref{eq::betadef}), (\ref{eq::betaqcd})]
\item the \msbar{}--\drbar{} relation through order $\alpha_s^2$
  [\eqn{eq::asMS2DR_2}]; due to \eqn{eq::aemudec} one can set
  $\alpha_e = \asMS{}$ at this order
\item decoupling through order $\alpha_s^2$ [Eqs.\,(\ref{eq::asdec})
  with $n_f=5$, \eqn{eq::zetas}]; no evanescent couplings appear at this
  order (for a ``common scale approach'').  Note that if the decoupling
  of the supersymmetric particles were performed in several steps,
  $\alpha_e$ would appear in the two-loop decoupling coefficients
\item $\beta(\alpha_s)$ through three loops in \susy{}
  [Eqs.\,(\ref{eq::betadef}), (\ref{eq::betasusy})]
\end{enumerate}
An advantage of this procedure as compared to a multi-scale approach is
that the RGEs are only one-dimensional (there is only one coupling
constant), and that for $\alpha_e$ one can apply
Eq.~(\ref{eq::couplings_full}) and
Eq.~(\ref{eq::aemudec}).

Let us remark that in principle it is possible to decouple the top quark
separately:
\begin{equation}
  \begin{split}
    \asMS{,(5)}(M_Z)
    \quad\stackrel{(i')}{\to}\quad \asMS{,(5)}(\mu_t)
    \quad\stackrel{(ii')}{\to}\quad \asMS{,(6)}(\mu_t)
    \quad\stackrel{(iii')}{\to}\quad \asMS{,(6)}(\mudec)
  \end{split}
  \label{eq::asrunmt}
\end{equation}
and continue with steps $(ii)$--$(iv)$ of \eqn{eq::asrundec}, but now
using $n_f=6$ in all the formulas. $\mu_t$ is an additional decoupling
scale to be chosen of the order of $m_t$. The only new ingredient needed
is the decoupling constant for going from five to six quark flavours in
the \msbar{} scheme in step $(ii')$. It has been evaluated through three
and four loops in Refs.\,\cite{Chetyrkin:1997un} 
and~\cite{Schroder:2005hy,Chetyrkin:2005ia}, 
respectively. In any case, for a mass spectrum
as given by the benchmark point
SPS1a$^\prime$~\cite{Aguilar-Saavedra:2005pw}, for example, the separate
decoupling of the top quark implies a numerically small effect. This can
also be established by comparing ``Scenario D'' and ``Scenario C'' in
Ref.~\cite{Harlander:2005wm}.

For a direct application of Eqs.\,(\ref{eq::asdec}), (\ref{eq::zetas}),
and (\ref{eq::asMS2DR_2}) to steps $(ii)$ and $(iii)$ of
\eqn{eq::asrundec}, these equations need to be inverted. 
In fact, it may be convenient for
practical phenomenological analyses to combine steps $(ii)$ and $(iii)$
of \eqn{eq::asrundec} into a single formula:
\begin{equation}
\begin{split}
\asDR{,(\rm full)} = \asMS{,(n_f)}&\bigg\{
  1 + \frac{\asMS{,(n_f)}}{\pi}\left( \frac{1}{4} - \zeta_{s1}^{(n_f)} \right)
  \\&
  + \left(\frac{\asMS{,(n_f)}}{\pi}\right)^2
  \left[ \frac{11}{8} - \frac{n_f}{12} - \frac{1}{2}\,\zeta_{s1}^{(n_f)}
    + 2\,(\zeta_{s1}^{(n_f)})^2
    - \zeta_{s2}^{(n_f)}\right]
\bigg\}\,,
\end{split}
\label{eq::asDRfull}
\end{equation}
where the scale $\mudec$ has been suppressed in the notation. The
coefficients $\zeta_{s1}^{(n_f)}$ and  $\zeta_{s2}^{(n_f)}$ are given in
\eqn{eq::zetas}.
The numerical deviation of Eq.~(\ref{eq::asDRfull}) from the 
two-step procedure is well below $0.2\%$.

\subsection{Numerical results}

\begin{figure}[t]
  \epsfig{figure=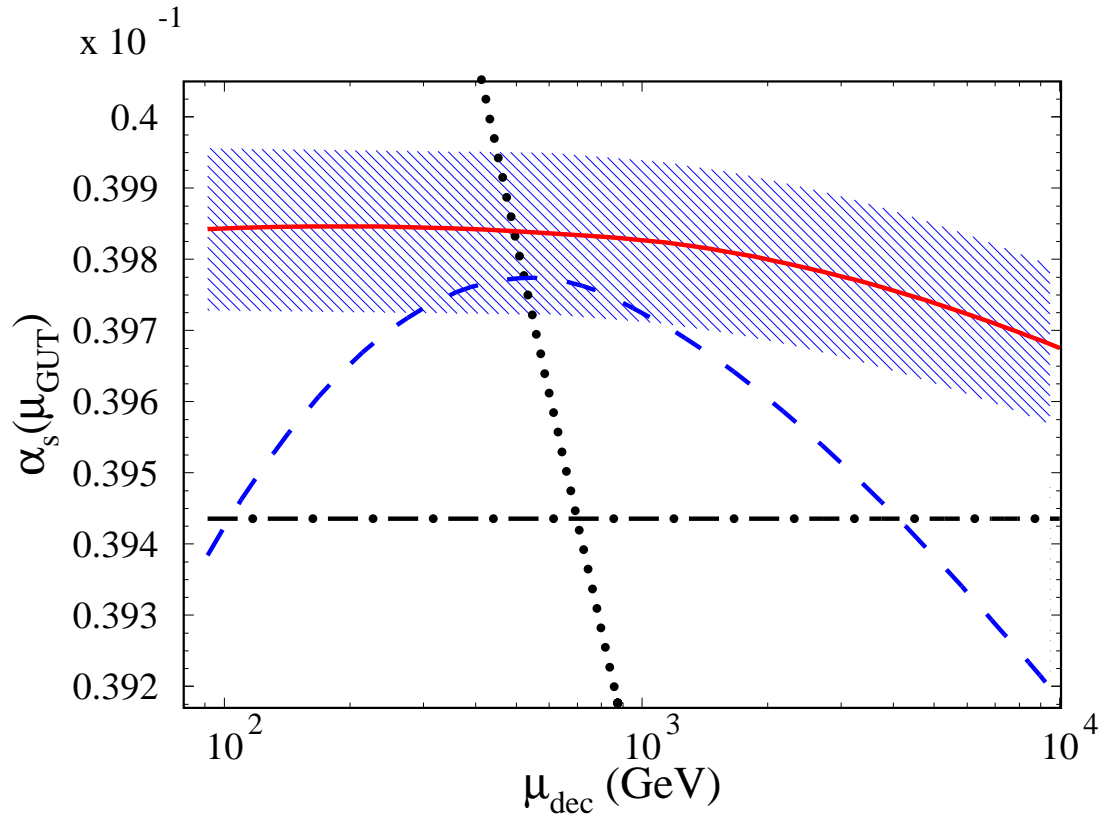,width=\textwidth}
  \caption{\label{fig::asGUT}
    $\alpha_s(\mu_{\rm GUT})$ as a function of $\mu_{\rm dec}$.
    Dotted, dashed and solid line: prescription discussed
    in Eq.~(\ref{eq::asrundec}) for one, two, and three loops, 
    respectively.
    The hatched band denotes the uncertainty from the input value
    $\alpha_s(M_Z)$. Dash-dotted line: 
    one-loop running and one-loop decoupling as described in
    Ref.~\cite{Aguilar-Saavedra:2005pw}.
  }
\end{figure}

The result for 
$\asDR{,(\rm full)}(\mugut=10^{16}~\mbox{GeV})$,
obtained using $M_Z = 91.1876~\mbox{GeV}$ and
\begin{equation}
  \begin{split}
    m_t=170.9\pm 1.9~\mbox{GeV}\,,\quad
    \asMS{,(5)}(M_Z) = 0.1189\,,\quad
    \tilde{M} = m_{\tilde{q}} = m_{\tilde{g}} = 1000\,\mbox{GeV}
  \end{split}
  \label{eq::num-input}
\end{equation}
as input parameters is shown if Fig.~\ref{fig::asGUT}.  The dotted,
dashed and solid line are based on \eqn{eq::asrundec}, where
$n$-loop running is combined with $(n-1)$-loop decoupling, as it is
required for consistency ($n=1,2,3$, respectively). 
We find a nice convergence when going from one to three
loops, with a very weakly $\mudec$--dependent result at three-loop order.

For comparison, we show the result obtained from the formula given in
Eq.~(21) of Ref~\cite{Aguilar-Saavedra:2005pw}. It corresponds to the
dash-dotted line in Fig.~\ref{fig::asGUT}. In this case, the evolution
is done in two steps:
\begin{equation}
\begin{split}
    \asMS{,(5)}(M_Z)\quad
    \stackrel{(i)}{\longrightarrow}\quad \asDR{,(\rm full)}(M_Z)\quad
    \stackrel{(ii)}{\longrightarrow}\quad \asDR{,(\rm full)}(\mugut)\,.
\end{split}
\end{equation}
In $(i)$, $\asMS{,(5)}(M_Z)$ is converted into $\asDR{,\rm (full)}(M_Z)$
by means of resummed one-loop contributions originating from both the
change of scheme (from \msbar{} to \drbar{}) and the decoupling of heavy
particles.  In step $(ii)$, the evolution from $M_Z$ to $\mugut$ is
performed using the one-loop MSSM RGE in the \drbar{} scheme ($\beta_0$
from \eqn{eq::betasusy}).  A remarkable feature of this combination is
that the $\mudec$--dependence drops out explicitely. However, the
difference between our three-loop result with two-loop decoupling (upper
solid line) and the one-loop formula given in
Ref.\,\cite{Aguilar-Saavedra:2005pw} exceeds the experimental
uncertainty by almost a factor of four for sensible values of
$\mu_{\rm dec}$.  This uncertainty is indicated by the
hatched band, derived from $\delta\alpha_s(M_Z)= \pm
0.001$~\cite{Bethke:2006ac}. The formulae of
Ref.\,\cite{Aguilar-Saavedra:2005pw} should therefore be taken only as
rough estimates; once precision studies are required, one should rely on
the consistent treatment of running, decoupling, and \msbar{}--\drbar{}
conversion, as it is outlined here.

\begin{figure}[ht]
  \epsfig{figure=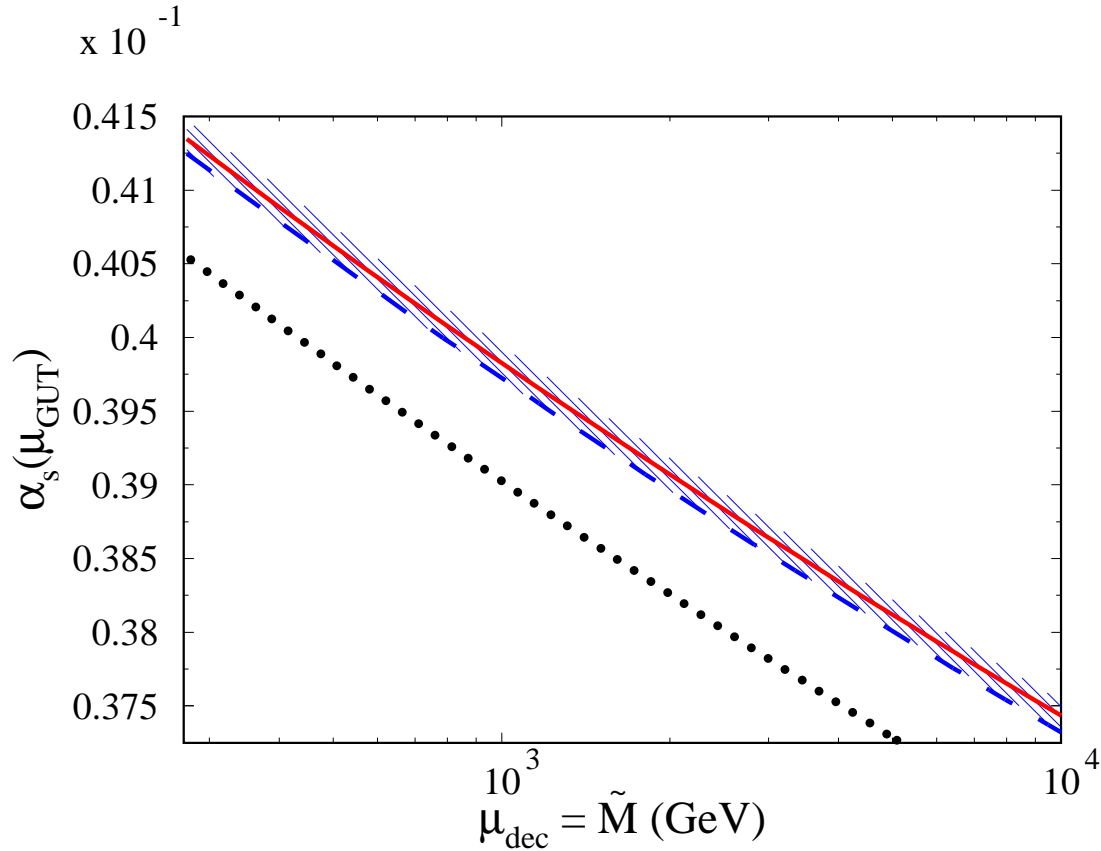,width=\textwidth}
  \caption{\label{fig::asGUTMtil}
    $\alpha_s(\mu_{\rm GUT})$ as a function of $\tilde{M}$
    where $\mu_{\rm dec}=\tilde{M}$ has been chosen.
    The notation is adopted from Fig.~\ref{fig::asGUT}.
  }
\end{figure}

In Fig.~\ref{fig::asGUTMtil} we show $\alpha_s(\mu_{\rm GUT})$
as a function of $\tilde{M}$ where $\mu_{\rm dec}=\tilde{M}$ has been
adopted. Dotted, dashed and full curve correspond again to the one-,
two- and three-loop analysis and the uncertainty form $\alpha_s(M_Z)$
is indicated by the hatched band. One observes a variation of
10\% as $\tilde{M}$ is varied between 
$100$~GeV and $10\,000$~GeV. This shows that the actual SUSY scale can
significantly influence the unification, respectively, the
non-unification behaviour of the three couplings 
at the GUT scale. 
For definite conclusions also the influence of $\tilde{M}$ 
on the electro-magnetic and the weak coupling
has to be carefully studied, of course.


\section{\label{sec::mb}Evaluating \boldmath{$m_b(\mu)$} in the
        \boldmath{$\drbarmath$} scheme}


Large values of $\tan\beta$ could enhance the importance of the bottom
Yukawa coupling in the MSSM enormously as compared to the SM. 
If predictions within this model are calculated in the \drbar{}
scheme, then it is certainly necessary to know $\mbDR{}$ with the
highest possible precision.  Experimental data concerning the
measurement of the bottom mass are typically converted into $\mu_b$
defined by the recursive equation
\begin{equation}
\begin{split}
\mu_b = m_b^{\msbarmath,(5)}(\mu_b)\,.
\end{split}
\end{equation}
In the following, we will provide a precise
relation between this value and $m_b^{\drbarmath,(\rm 5)}(\mu_S)$, where
$\mu_S$ may be an energy scale between the electroweak gauge boson
masses and a few TeV.

The quantity $\mbDR{,(5)}(M_Z)$ has also been considered in
Ref.\,\cite{Baer:2002ek} at two-loop level. We will comment on the
differences between this and our result at the end of this section.


\subsection{Evolution and scheme conversion of {\boldmath $m_b^{(5)}$}}
We follow two different methods. The first one is given by the chain
\begin{equation}
\begin{split}
m_b^{\msbarmath,(5)}(\mu_b)\quad\stackrel{(i)}{\to}\quad
m_b^{\msbarmath,(5)}(\mu_S)\quad\stackrel{(ii)}{\to}\quad
m_b^{\overline{\rm DR},(5)}(\mu_S)\,.
\label{eq::mbrun}
\end{split}
\end{equation}
The RGEs of QCD required for step $(i)$ are known to four-loop
accuracy~\cite{Chetyrkin:1997dh,Vermaseren:1997fq}, and three-loop
corrections are available~\cite{Harlander:2006rj} for the
\msbar{}--\drbar{} conversion relation in step $(ii)$. A peculiar
feature of this mass conversion is that the evanescent coupling
$\alpha_e$ occurs already at one-loop level:
\begin{eqnarray}
  \mDRbar &=& \mMSbar\left(1 -\frac{\alpha_e}{3\pi}+ \ldots \right)
  \,,
  \label{eq::mMS2DR_2}
\end{eqnarray}
where the dots denote higher orders in $\alpha_e$, $\eta_r$ and $\alpha_s$.

The second method consists of the following sequence:
\begin{equation}
\begin{split}
m_b^{\msbarmath,(5)}(\mu_b)\quad\stackrel{(i')}{\to}\quad
m_b^{\drbarmath,(5)}(\mu_b)\quad\stackrel{(ii')}{\to}\quad
m_b^{\overline{\rm DR},(5)}(\mu_S)\,.
\label{eq::mbrunprime}
\end{split}
\end{equation}
Again, three-loop terms are known for the relation used in step
($i^\prime$). The running in the \drbar{} scheme in step ($ii^\prime$)
involves the non-diagonal RGEs of $\asDR{,(5)}$ and the evanescent
couplings $\alpha_e^{(5)}$ and $\eta_r^{(5)}$. The corresponding $\beta$
functions are known to four-, three-, and one-loop order,
respectively~\cite{Harlander:2006rj}.

Because of these multi-dimensional RGEs required for
\eqn{eq::mbrunprime}, the sequence of \eqn{eq::mbrun} is preferable to
the latter in practical calculations. However, it may be interesting to
see how the perturbative series behaves for the two methods, and to
which extent they are numerically equivalent.  In particular, this may
provide an estimate of the uncertainty in the relation between
$m_b^{\overline{\rm MS},(5)}(\mu_b)$ and $m_b^{\overline{\rm
DR},(5)}(\mu_S)$.

Just as for the strong coupling $\alpha_s$, the consistent evolution of
$m_b$ requires that $n$-loop running is combined with $(n-1)$-loop
\msbar{}--\drbar{} conversion. 
Assuming that the couplings $\asMS{,(5)}(M_Z)$, $\asDR{,(5)}(M_Z)$,
$\alpha_e^{(5)}(M_Z)$ and $\eta_r^{(5)}(M_Z)$ are known, all operations
are within five-flavour QCD and do not require any
decoupling. This allows it to apply \eqn{eq::mbrun} at four-loop order.
This is not strictly the case for \eqn{eq::mbrunprime}, because the
$\beta$ functions for the evanescent couplings are not known to four
loops.  We ignore this fact when evaluating the three- and four-loop
evolution along \eqn{eq::mbrunprime}, expecting these effects to be small.


\subsection{Input values for {\boldmath $\alpha_e$} and {\boldmath $\eta_r$}}
\label{sec::aeetar}
In addition to $m_b^{\msbarmath,(5)}(\mu_b)$, the only input value for the
determination of $m_b^{\drbarmath,(5)}(\mu_S)$ is $\asMS{,(5)}(M_Z)$.
However, either of the two methods of Eqs.\,(\ref{eq::mbrun}) and
(\ref{eq::mbrunprime}) requires to switch to the \drbar{} parameters
$\asDR{,(5)}$, $\alpha_e^{(5)}$, and $\eta_r^{(5)}$ at some point.

In principle, any choice of $\asDR{,(5)}(M_Z)$, $\alpha_e^{(5)}(M_Z)$
and $\eta_r^{(5)}(M_Z)$ is allowed
which obeys Eq.~(\ref{eq::asMS2DR_2}) for the given value of
$\asMS{,(5)}(M_Z)$. Any such choice would simply correspond to a
particular renormalization scheme.  But let us assume that QCD is the
low energy effective theory of SUSY-QCD, and thus
\eqn{eq::couplings_full} holds. 
Then all $\overline{\rm DR}$ couplings of five-flavour standard QCD are 
uniquely determined by the corresponding decoupling relations.
In contrast to Section\,\ref{sec::as}, we
now want to go to the four-loop level, and thus we take the
one-loop decoupling relation for $\alpha_e$ into
account\footnote{Since we take the three-loop beta function for
  $\alpha_e$ into account we would need the two-loop corrections at
  this point. However, we assume that the numerical effect is small.}, 
see~\eqn{eq::decoupling_ev}. The $\eta_r$, on the other hand, occur only at
four-loop level and can be decoupled trivially by setting $\eta_r^{\rm
(full)}(\mudec) = \eta_r^{\rm (5)}(\mudec)$.

Since $\asDR{,(\rm full)}$ is not known {\it a priori}, one cannot use
\eqn{eq::decoupling_ev} directly in order to derive
$\alpha_e^{(5)}$. Rather, we start with a trial value for $\asDR{,(\rm
full)}(\mudec)$ and obtain the corresponding $\alpha_e^{(5)}(\mudec)$
through \eqn{eq::decoupling_ev}, as well as $\asDR{,(5)}(\mudec)$
through \eqn{eq::zetas}.  Then we evaluate $\asMS{,(5)}(\mudec)$ through
\eqn{eq::asMS2DR_2}, and from that $\asMS{,(5)}(M_Z)$.  The trial value
for $\asDR{,(\rm full)}(\mudec)$ is systematically varied until the
resulting $\asMS{,(5)}(M_Z)$ agrees with the experimental input.

As indicated above, current knowledge of the renormalization group
functions allows us to follow the sequences \eqn{eq::mbrun} and
\eqn{eq::mbrunprime} at four-loop level. Strictly speaking, this
requires also that $\asDR{,(5)}(M_Z)$, $\alpha_e^{(5)}(M_Z)$, and
$\eta_r^{(5)}(M_Z)$ are derived from $\asMS{,(5)}(M_Z)$ at four-loop
level. The decoupling of the SUSY particles mentioned above should thus
be performed at three-loop level, which is currently not known for any
of the three couplings.  However, we may consider $\asDR{,(5)}(M_Z)$,
$\alpha_e^{(5)}(M_Z)$, and $\eta_r^{(5)}(M_Z)$ as input, and their
derivation from $\asMS{,(5)}(M_Z)$ only as a guideline, allowing us to
neglect higher order decoupling effects.  We do not expect that a fully
consistent evaluation of these parameters will change our numerical
results for $\mbDR{}$ significantly.


\subsection{Numerical results}

Taking Eq.~(\ref{eq::num-input}) as well as
$m_b^{\overline{\rm MS},(5)}(\mu_b)=4.164$~GeV~\cite{Kuhn:2007vp} 
as input values and following steps $(i)$
and $(ii)$ of \eqn{eq::mbrun}, we find (the numbers refer to $\mu_{\rm
  dec} = \mu_S = M_Z$):
\begin{itemize}
\item[($i$)] The difference between two- and three-loop running on
  $m_b^{\overline{\rm DR},(5)}(M_Z)$ is 23~MeV, while going to four
  loops further modifies the result by less than 2~MeV.
\item[($ii$)] The one-loop term in the \msbar{}--\drbar{} transition
amounts for $\mu=M_Z$
to about 29~MeV, while the two-loop term is only about 2~MeV.
\end{itemize}

Following steps $(i^\prime)$ and $(ii^\prime)$ of \eqn{eq::mbrunprime},
on the other hand, one finds:
\begin{itemize}
\item[$(i^\prime)$] The one-loop term in the \msbar{}--\drbar{}
transition amounts for $\mu=\mu_b$ to about 66~MeV, while the two-loop
term is only about 6~MeV.
\item[$(ii^\prime)$]The difference between two- and three-loop running on
  $m_b^{\overline{\rm DR},(5)}(M_Z)$ is 18~MeV, while going to four
  loops further modifies the result by less than 0.5~MeV.
\end{itemize}

\begin{figure}[t]
  \begin{center}
    \begin{tabular}{c}
      \leavevmode
      \epsfxsize=\textwidth
      \epsffile{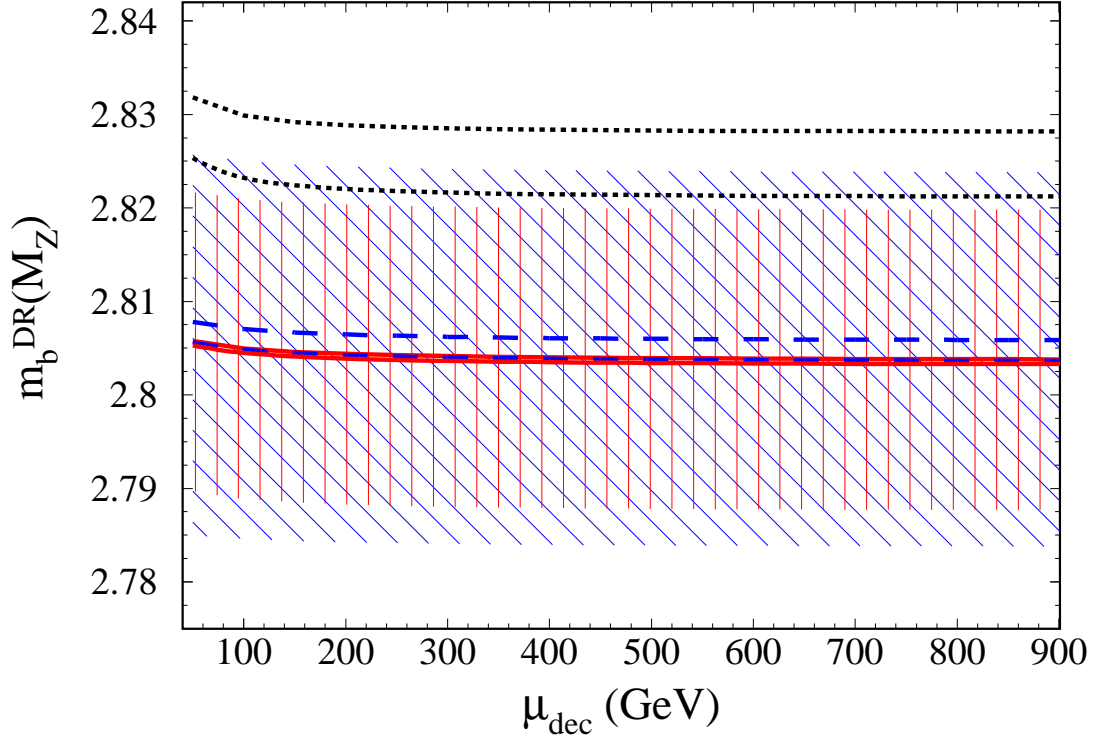}
    \end{tabular}
  \end{center}
  \caption{\label{fig::mb_mudec}
    $m_b^{\overline{\rm DR}}(M_Z)$ as a function of $\mudec$
    for two-, three- and four-loop running (dotted, dashed and solid
    lines). The upper (lower) curves correspond to the prescriptions
    ($i$) and ($ii$) (($i^\prime$) and ($ii^\prime$)).
    The band describes the uncertainty from $\alpha_s$ or $m_b$ (see text).
  }
\end{figure}

In Fig.~\ref{fig::mb_mudec}, $m_b^{\overline{\rm DR}}(M_Z)$ is shown as
a function of $\mudec$ where the dotted, dashed and solid lines
correspond to two-, three- and four-loop 
running.\footnote{For the one-loop result the difference between
  $(i),(ii)$ and $(i^\prime),(ii^\prime)$ is small because
  no non-trivial matching terms are included at this order.
  The resulting value $m_b^{\overline{\rm DR}}(M_Z)\approx 3.05$~GeV lies
  well outside the plot.
}
At each loop level,
the upper curve corresponds to the path given in \eqn{eq::mbrun}, the
lower one to \eqn{eq::mbrunprime}.  The smaller hatched band reflects the
uncertainty induced by the current experimental accuracy of
$\asMS{,(5)}(M_Z)$, equal to $\delta
\alpha_s=0.001$~\cite{Bethke:2006ac},
and the larger one corresponds to $\delta
m_b=25$~MeV~\cite{Kuhn:2007vp}.
The total uncertainty from both $m_b$ and $\alpha_s$ is thus obtained by
adding in quadrature the individual  uncertainties.

As expected, the difference between paths \eqn{eq::mbrun} and
\eqn{eq::mbrunprime} decreases as higher order corrections are
included. At four-loop level, the lines are practically
indistinguishable. One notices a slightly better convergence of the
perturbative terms when following the path of \eqn{eq::mbrunprime}.  Note also
that both two-loop curves lie on the upper edge or even outside
the uncertainty band of $\alpha_s$
and $m_b$, thus proving the importance of the three-loop terms.
Detailed numerical results for the couplings and bottom quark masses are
given in Tab.~\ref{tab::res1}.  We consider the choices $\mu_S = \mudec
= M_Z$, as well as $\mu_S = \mudec/2 = M_Z$.

\begin{table}[t]
{\scalefont{1.0}
\begin{center}
\begin{tabular}{l|llll}
\hline
\multicolumn{5}{c}{$\mu_S=\mu_{\rm dec}=M_Z$} \\
\hline
($i$), ($ii$) & 1 & 2 & 3 & 4 (running)
\\\hline
$\alpha_e^{(5)}(\mu_S)$
 &  0.1189 &  0.0968 &  0.0995 &  0.0995
\\$\alpha_s^{\overline{\rm DR},(5)}(\mu_S)$ 
 &  0.1189 &  0.1200 &  0.1202 &  0.1202
\\
$m_b^{\overline{\rm MS},(5)}(\mu_S)$
 &   3.055 &   2.859 &   2.838 &   2.836\\
$m_b^{\overline{\rm DR},(5)}(\mu_S)$
 &   3.055 &   2.830 &   2.807 &   2.805\\
\hline
($i^\prime$), ($ii^\prime$)
\\\hline
$\alpha_e^{(5)}(\mu_b)$ 
 &  0.1649 &  0.1501 &  0.1537 &  0.1538
\\$\alpha_s^{\overline{\rm DR},(5)}(\mu_b)$
 &  0.2153 &  0.2316 &  0.2336 &  0.2343
\\$m_b^{\overline{\rm DR},(5)}(\mu_b)$
 &   4.164 &   4.098 &   4.092 &   4.094
\\$m_b^{\overline{\rm DR},(5)}(\mu_S)$
 &   3.055 &   2.823 &   2.805 &   2.805\\
\hline
\multicolumn{5}{c}{} \\[-.5em]
\hline
\multicolumn{5}{c}{$2\mu_S=\mu_{\rm dec}=2 M_Z$} \\
\hline
($i$), ($ii$) & 1 & 2 & 3 & 4 (running)
\\\hline
$\alpha_e^{(5)}(\mu_S)$
 &  0.1133 &  0.1006 &  0.1014 &  0.1014
\\$\alpha_s^{\overline{\rm DR},(5)}(\mu_S)$ 
 & 0.1189 &  0.1200 &  0.1202 &  0.1202
\\$m_b^{\overline{\rm MS},(5)}(\mu_S)$
 &  3.055 &   2.859 &   2.838 &   2.836
\\$m_b^{\overline{\rm DR},(5)}(\mu_S)$
 &   3.055 &   2.829 &   2.807 &   2.804 \\
\hline
($i\prime$), ($ii^\prime$)
\\\hline
$\alpha_e^{(5)}(\mu_b)$ 
 &  0.1590 &  0.1545 &  0.1559 &  0.1560
\\$\alpha_s^{\overline{\rm DR},(5)}(\mu_b)$
 &  0.2153 &  0.2316 &  0.2336 &  0.2343
\\$m_b^{\overline{\rm DR},(5)}(\mu_b)$
 &   4.164 &   4.096 &   4.091 &   4.092
\\$m_b^{\overline{\rm DR},(5)}(\mu_S)$
 &   3.055 &   2.822 &   2.804 &   2.804\\
\hline
\end{tabular}
  \caption{\label{tab::res1}
    Numerical results for the bottom quark \drbar{} mass for
    $\mu_S=M_Z$ using $\mu_{\rm dec}=M_Z$ and $\mu_{\rm dec}=2M_Z$,
    respectively. For convenience intermediate results for
    $\alpha_s$ and $\alpha_e$ are given resulting from the one-, two-,
    three- and four-loop analysis. 
    }
\end{center}
}
\end{table}

It is worth mentioning that the numerical effect of the decoupling of
$\asDR{,(n_f)}$ (cf.\ Eq.~(\ref{eq::asdec})) and $\alpha_e$
(cf.\ Eq.~(\ref{eq::decoupling_ev})) on
$m_b^{\drbarmath,(5)}(M_Z)$ is not negligible.  For $\zeta_s^{(n_f)}=1$
and  $\zeta_e=1$, we observe a decrease of more than $6$~MeV
for both Eqs.\,(\ref{eq::mbrun})
and (\ref{eq::mbrunprime}), employed at  two-, three-, and four-loop
order.

\begin{figure}[ht]
  \epsfig{figure=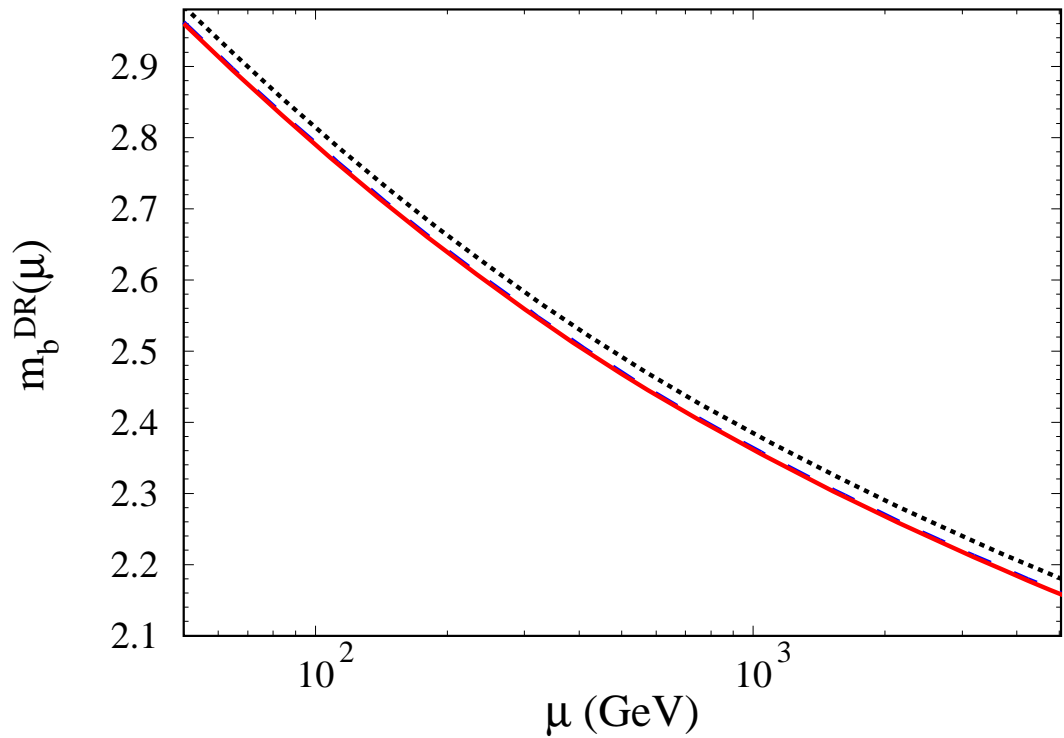,width=\textwidth}
  \caption{\label{fig::mbmu}
    $m_b^{\rm DR}(\mu)$ as a function of $\mu$.
    The notation is adopted from Fig.~\ref{fig::mb_mudec}.
  }
\end{figure}

In Fig.~\ref{fig::mbmu} we show $\mbDR{}(\mu)$ varying $\mu$ between
50~GeV and 5~TeV where the two-, three- and four-loop results
corresponding to the steps $(i)$ and $(ii)$ of Eq.~(\ref{eq::mbrun})
are plotted. The decoupling scale is set to $\mu_{\rm dec}=\tilde{M}=1$~TeV.
In Tab.~\ref{tab::mbDRmu} we show for some selected
values of $\mu$ the corresponding result for $\mbDR{}$ together with the
uncertainties arising from $\delta\alpha_s^{\overline{\rm MS}}=0.001$
and $\delta m_b^{\overline{\rm MS}}(\mu_b)=25$~MeV.\footnote{In 
  the analysis of
  Ref.~\cite{Kuhn:2007vp} $\delta\alpha_s^{\overline{\rm MS}}=0.002$ has been
  adopted, leading to the uncertainty of 25~MeV for $m_b$. Still, for
  illustration purpose we choose  
  $\delta\alpha_s^{\overline{\rm MS}}=0.001$ in order to obtain the numbers in
  Tab.~\ref{tab::mbDRmu}.}

\begin{table}
  \begin{center}
    \begin{tabular}{c|l}
      $\mu$ (GeV) & $\mbDR{}(\mu)$ \\
      \hline
91.1876 & $  2.804(16)(20)$ \\
350 & $  2.528(17)(18)$ \\
500 & $  2.467(17)(18)$ \\
800 & $  2.394(17)(17)$ \\
1000 & $  2.361(17)(17)$ \\
2000 & $  2.268(17)(16)$ \\
    \end{tabular}
    \caption{\label{tab::mbDRmu}
      $\mbDR{}$ for various values of the renormalization scale.
      The two numbers given in the round brackets correspond to
      the uncertainties induced by $\delta\alpha_s^{\overline{\rm
      MS}}=0.001$ (first
      number) and $\delta m_b^{\overline{\rm MS}}(\mu_b)=25$~MeV 
      (second number), 
      respectively.}
  \end{center}
\end{table}

Let us now comment on the earlier two-loop calculation of $\mbDR{}$ by
Baer et al.~\cite{Baer:2002ek}. Our results represent an improvement
with respect to several issues. First of all, we increased the accuracy
of the result by including the three- and the four-loop terms in the
derivation of $\mbDR{}$. In particular the three-loop terms turn out to
be numerically very important, while the four-loop result indicates a
nice stabilization of the perturbative expansion.  In addition, we have
made a significant conceptual generalization, related to the evanescent
couplings.  Ref.\,\cite{Baer:2002ek} sets $\alpha_s(M_Z)=\alpha_e(M_Z)$
in \eqn{eq::mMS2DR_2}. As we have argued above, this corresponds to a
particular renormalization scheme within \drbar{}, and the value for
$\mbDR{}$ is associated with this scheme. In contrast to that, we have
systematically derived $\alpha_e$ by assuming that squarks and gluinos
decouple at a scale $\mudec$ and have studied the dependence of
$\mbDR{}$ on this scale.  Finally, we have given explicit numerical
results for the method described by
\eqn{eq::mbrunprime}. Ref.\,\cite{Baer:2002ek} correctly claims that the
difference to the method of \eqn{eq::mbrun} is small, but no numbers or
any details of the calculation are given. In particular, it is unclear
which value was used for $\alpha_e(\mu_b)$ in \eqn{eq::mMS2DR_2}.

It should be pointed out that the formulae and prescriptions of
Ref.~\cite{Baer:2002ek} have been partially taken over in the outline of
the SPA project~\cite{Aguilar-Saavedra:2005pw}. However, as in the case
of the strong coupling, various orders of perturbation theory have been
combined inconsistently. Thus, let us stress again that the formulae
of Ref.~\cite{Aguilar-Saavedra:2005pw} should not be taken over
literally in phenomenological analyses if precision is of concern.



\section{Conclusions}

Application of \dred{} to non-SUSY theories is rather cumbersome
compared to \dreg{} because of the occurrence of evanescent couplings.
However, when low energy precision data such as 
$\alpha_s^{\overline{\rm MS}}(M_Z)$ or
$m_b^{\overline{\rm MS}}(\mu_b)$ are 
related to their counterparts in a SUSY theory at high
energies, one may need to switch between the \msbar{} and the \drbar{}
scheme at some energy scale. The conversion formulae will involve
evanescent couplings, sometimes already at one-loop level, as in the
case of the quark mass.

We have used recent three- and four-loop results for the $\beta$
functions, the quark anomalous dimension, and the decoupling
coefficients in order to derive $\alpha_s^{\overline{\rm DR}}(\mugut)$ and
$\mbDR{,(5)}(\mu)$ from $\asMS{}(M_Z)$ and $m_b^{\overline{\rm
    MS}}(\mu_b)$ 
at three- and
four-loop level, respectively.

It turns out that the three-loop terms are numerically significant both
for $\alpha_s$ and for $m_b$. The dependence on where the SUSY spectrum
is decoupled becomes particularly flat in this case.
The theoretical uncertainty is expected to be negligible w.r.t.\ the
uncertainty induced by the experimental input values.

Comparing our results and methods to the literature, we find that the
issue of evanescent couplings has either been ignored (by assuming
$\alpha_e=\alpha_s$) or circumvented by decoupling the SUSY spectrum at
$\mudec=M_Z$. We find that at one- and two-loop level, this choice does
not allow for a good approximation of the higher order effects, if one
assumes the SUSY partner masses to be of the order of 1\,TeV.

To conclude, we strongly suggest that phenomenological studies
concerning the implications of low energy data on Grand Unification
should be done at three-loop level, and that decoupling effects and
evanescent couplings are properly taken into account.


\bigskip
\noindent
{\large\bf Acknowledgements}\\ 
This work was supported by the DFG through SFB/TR~9 and HA 2990/3-1.



\end{document}